\documentclass[10pt,twocolumn]{article} 
\usepackage{simpleConference}
\usepackage{times}
\usepackage{graphicx}
\usepackage{amssymb}
\usepackage{amsfonts}
\usepackage{amsmath}
\usepackage{placeins}
\usepackage{booktabs}
\usepackage{url,hyperref}
\usepackage{multirow}
\usepackage{multicol}
\usepackage{float}
\usepackage{siunitx}
\usepackage{tikz}
\usepackage{subcaption}
\usepackage{hyperref}  
\newcommand{\figref}[2]{\hyperref[#1]{\ref*{#1}#2}}

\newcommand{\labelB}{{\tt B}}
\newcommand{\labelA}{{\tt A}}

\begin{document}

\title{An Analytical Neighborhood Enrichment Score for Spatial Omics}

\author{
Axel Andersson\textsuperscript{1,*} and Hanna Nyström\textsuperscript{1} \\
\textsuperscript{1}Department of Diagnostics and Intervention, Surgery \\
Umeå University, Umeå, Sweden \\
\textsuperscript{*}Corresponding author: axel.andersson@umu.se
}
\maketitle
\thispagestyle{empty}

\begin{abstract}
\noindent The neighborhood enrichment test is used to quantify spatial enrichment and depletion between spatial points with categorical labels, which is a common data type in spatial omics. Traditionally, this test relies on a permutation-based Monte Carlo approach, which tends to be computationally expensive for large datasets. In this study, we present a modified version of the test that can be computed analytically. This analytical version showed a minimum Pearson correlation of 0.95 with the conventional Monte Carlo-based method across eight spatial omics datasets, but with substantial speed-ups. Additional experiments on a large Xenium dataset demonstrated the method's ability to efficiently analyze large-scale data, making it a valuable tool for analyzing spatial omics data.
\end{abstract}

\section{Introduction}
\noindent Spatial omics, the study of the spatial organization of biomolecules, is a rapidly growing field~\cite{bressan2023dawn}. Many spatial omics techniques produce image data that, after analysis, yield spatial points with categorical labels. For example, these points may represent locations of different cell types or mRNA molecules.

The neighborhood enrichment test~\cite{squidpy} is a widely used permutation-based method to quantify spatial enrichment or depletion between points with different labels. It is implemented in several statistical toolboxes for spatial omics~\cite{squidpy,schapiro2017histocat,behanova2023visualization,crawldad} and has been applied in numerous studies~\cite{example1,example2}. The test measures how often spatial points with label {\tt A} are neighbors to those with label {\tt B}, then compares this observed count to a null distribution generated by repeatedly shuffling the labels across the data set. The output is a z-score-like value that indicates whether the two labels are spatially enriched or depleted. However, the repeated shuffles required to estimate this distribution can be computationally intensive, particularly for large datasets with many labels. We therefore propose a modified version of the enrichment score that can be computed analytically. This approach significantly improves computational speed and scalability. The code is publicly available at \url{https://github.com/wahlby-lab/analytical-enrichment-test}.

\section{Materials and methods}    
\subsection*{The neighborhood enrichment score}
\noindent The enrichment score, as implemented in the popular spatial omics toolbox Squidpy~\cite{squidpy}, is given by:
\begin{equation}
    z_{\text{{\tt AB}}} = \frac{x_{\text{{\tt AB}}} - \mu_{\text{{\tt AB}}}}{\sigma_{\text{{\tt AB}}}}
    \label{eq:1}
\end{equation}
where \(x_{\text{{\tt AB}}}\) is the count of points with label {\tt A}  that are neighbors points with label {\tt B}. A neighbor can be defined in different ways, for example, if two points are within a set distance or if a point is within the $k$-nearest neighbors of a reference point. The variables \(\mu_{\text{{\tt AB}}}\) and \(\sigma_{\text{{\tt AB}}}\) in Eq.~\ref{eq:1} correspond to the expectation and standard deviation in the number of neighbors between {\tt A}-labeled points and {\tt B}-labeled points after shuffling the labels of the points. The number of shuffles, $N_{\text{MC}}$, used to estimate \(\mu_{\text{{\tt AB}}}\) and \(\sigma_{\text{{\tt AB}}}\) is given as a user-defined parameter and should be large to avoid a high variance in the estimated z scores. Default in Squidpy is $N_{\text{MC}} = 10^3$~\cite{squidpy}.

\subsection*{Analytical neighborhood enrichment score}
\noindent Instead of shuffling the labels and estimating the mean and standard deviation of neighbor counts through Monte Carlo sampling, we propose an analytical alternative that can be computed directly. Consider two labels: a reference label \labelA, and a neighbor label \labelB, and define:
\begin{itemize}
    \item $\mathbf{b} \in \{0,1\}^N$ : a binary vector indicating which of the $N$ points have label \labelB,
    \item $\mathbf{W} \in \{0, 1\}^{N \times N}$ : a sparse adjacency matrix encoding the neighborhood graph,
    \item $\mathbf{1} \in \mathbb{R}^N$ : a vector of ones (i.e., $\mathbf{1}_i = 1$ for all $i$).
\end{itemize}
The vector containing the number of \labelB-labeled neighbors for each point is:
\begin{equation*}
    \mathbf{y} = \mathbf{W} \mathbf{b}.
\end{equation*}
The expected number of \labelB-labeled neighbors for a randomly selected point is:
\begin{equation*}
    \mathbb{E}\big[\mathbf{y}\big] = \frac{1}{N} \mathbf{1}^\top \mathbf{y},
\end{equation*}
and the variance is:
\begin{equation*}
    \mathbb{V}\big[\mathbf{y}\big] = \mathbb{E}\big[\mathbf{y}^{\circ 2}\big] - \mathbb{E}\big[\mathbf{y}\big]^2,
\end{equation*}
where $(\cdot)^{\circ 2}$ refers to the Hadamard square. Let $n_{\text{\labelA}}$ be the number of points with the label \labelA. If we sample $n_{\text{\labelA}}$ points independently with replacement, the expected total number of \labelB-labeled neighbors is:
\begin{equation*}
    \mu_{\text{\labelA\labelB}} = n_{\text{\labelA}} \cdot  \mathbb{E}\big[\mathbf{y}\big],
\end{equation*}
with variance:
\begin{equation*}
    \sigma_{\text{\labelA\labelB}}^2 =  n_{\text{\labelA}} \cdot \mathbb{V}\big[\mathbf{y}\big].
\end{equation*}
The proposed analytical enrichment z-score is then given by:

\begin{equation}
    z_{\text{\labelA\labelB}} = \frac{o_{\text{\labelA\labelB}} - \mu_{\text{\labelA\labelB}}}{\sigma_{\text{\labelA\labelB}}},
    \label{Z}
\end{equation}
where $o_{\text{\labelA\labelB}}$ is the total number of neighbors labeled \labelB{} of points labeled \labelA, computed as:
\begin{equation*}
    o_{\text{\labelA\labelB}} = \mathbf{a}^\top \mathbf{y}.
\end{equation*}
with $\mathbf{a} \in \{0, 1\}^N$ as the indicator vector for label \labelA{} points. Equation~\ref{Z} can be rewritten as:
\begin{equation*}
    z_{\text{\labelA\labelB}} = \sqrt{n_{\text{\labelA}}} \cdot \frac{\bar{o}_{\text{\labelA\labelB}} - \mathbb{E}[\mathbf{y}]}{\sqrt{\mathbb{V}[\mathbf{y}]}}
\end{equation*}
where  $\bar{o}_{\text{\labelA\labelB}} = \frac{o_{\text{\labelA\labelB}}}{n_{\text{\labelA}}}$.

\subsection*{Matrix formulation of the enrichment score}
\noindent
The previous pairwise formulation can be extended to a matrix-based form that simultaneously computes enrichment scores for all label pairs using basic linear algebra. Define:
\begin{itemize}
    \item $\mathbf{L} \in \{0,1\}^{N \times K}$: a one-hot label matrix, where $L_{ik} = 1$ if the point $i$ has the label $k$ and $K$ is the number of unique labels.
    \item $\mathbf{N} = \mathrm{diag} \big(\mathbf{L}^\top \mathbf{1}\big) \in \mathbb{N}^{K\times K}$ : a diagonal matrix where $N_{kk}$ is the number of points with the label $k$.
\end{itemize}
The matrix of neighbor label counts is given by:
\begin{equation}
    \mathbf{Y} = \mathbf{W} \mathbf{L} \in \mathbb{R}^{N \times K},
\end{equation}
where each column $\mathbf{Y}_{:,k}$ contains the number of neighbors with label $k$ for each point. The observed number of neighbors labeled $k$ for each point labeled $j$ is given by the matrix:
\begin{equation*}
    \mathbf{O} = \mathbf{L}^\top \mathbf{Y} = \mathbf{L}^\top \mathbf{W} \mathbf{L} \in \mathbb{R}^{K \times K}
\end{equation*}
and the corresponding per-point average is:
\begin{equation*}
\bar{\mathbf{O}} = \mathbf{N}^{-1} \mathbf{O}.
\end{equation*}
 The matrix of enrichment z-scores between all pairs of labels is:
\begin{equation*}
    \mathbf{Z} = \mathbf{N}^{1/2} \left(\frac{\bar{\mathbf{O}} - \mathbb{E}\big[\mathbf{Y}\big]}{\sqrt{\mathbb{V}\big[\mathbf{Y}\big]}}\right) \in \mathbb{R}^{K \times K},
\end{equation*}
where all square roots, divisions, and differences are applied element-wise. Here, $\mathbb{E}[\mathbf{Y}] \in \mathbb{R}^{1 \times K}$ and $\mathbb{V}[\mathbf{Y}] \in \mathbb{R}^{1 \times K}$ are the expected values and variances of $\mathbf{Y}$ across all points, calculated column-wise.
\medskip

\begin{table}
\centering
\caption{Dataset characteristics including the number of unique labels and total number of points.}
\begin{tabular}{lllr}
\hline
\textbf{Dataset} & \textbf{\shortstack{\# of labels}} & \textbf{\# of points} & \textbf{Ref.} \\
\hline
MIBI-TOF & 8 & 1.2k & \cite{mibitof}\\
4i & 15 & 2.8k & \cite{four_i}\\
Visium & 12 & 2.8k &  \cite{squidpy_visium} \\
IMC & 11 & 4.7k & \cite{IMC}\\
MERFISH I & 16 & 6.5k & \cite{merfish}\\
SeqFISH & 22 & 19.4k & \cite{seqfish}\\
osmFISH & 36 & 1.98M & \cite{Codeluppi2018}\\
MERFISH II & 135 & 3.73M & \cite{merfish} \\\hline
\end{tabular}
\label{table:datasets}
\end{table}

\begin{figure*}[h!]
    \centering
    \includegraphics[trim=0 0cm 0 0,clip, width=\textwidth]{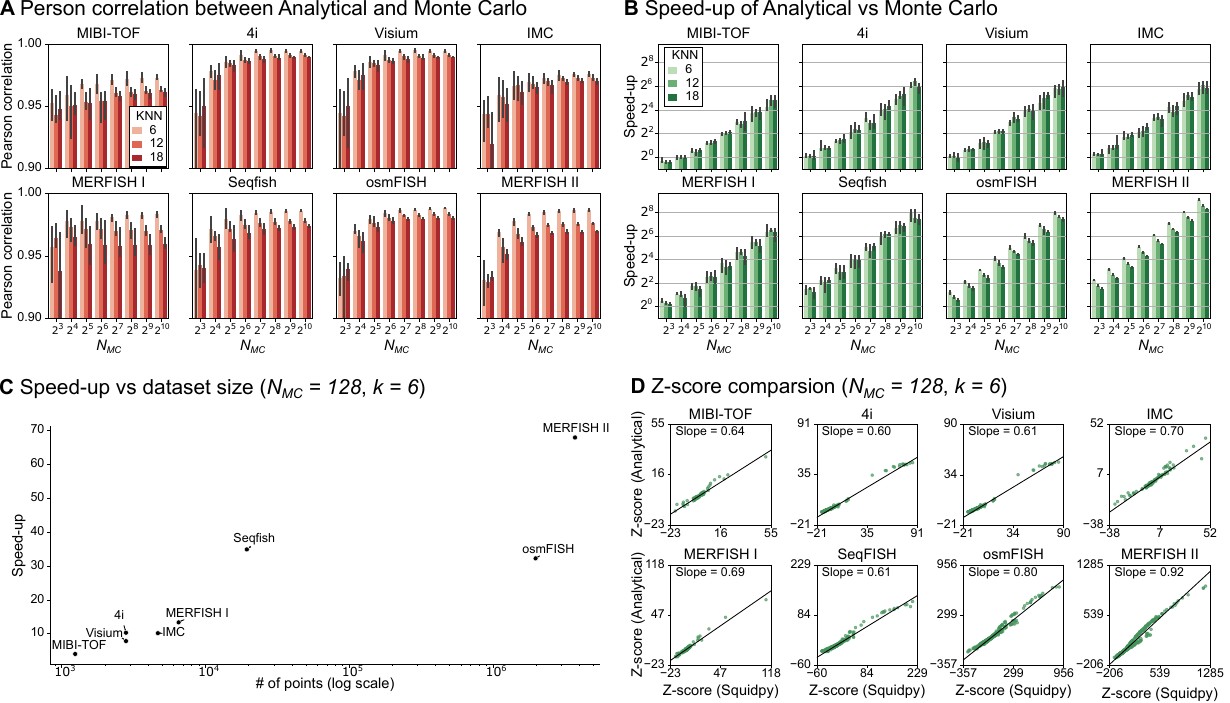}
    \caption{The analytical enrichment score shows strong agreement with traditional Monte Carlo-based method across multiple datasets, while providing substantial speed-up. \textbf{\textsf{A}}~Pearson correlation between the proposed {Analytical} enrichment score and the Monte Carlo-based implementation in Squidpy~\cite{squidpy} with different number of label-permutations ($N_{\mathrm{MC}}$).
    \textbf{\textsf{B}}~Speed-up of the analytical method compared to the Monte Carlo-based approach.  Experiments were conducted using neighborhood graphs constructed with $k = 6$, $k = 12$, or $k = 18$ nearest neighbors. \textbf{\textsf{C}} Speed-up of the analytical method compared to the Monte Carlo-based method for different dataset sizes (number of points). Speed-up increased with dataset size. \textbf{\textsf{D}}~Comparison of z-scores computed by the analytical and Monte Carlo-based methods. Each point represents the enrichment z-score for a pair of labels in a given dataset. Overall, the z-scores produced by the analytical methods are lower than those of the Monte Carlo-based method. }
    \label{fig:results}
\end{figure*}

\section{Experiments and results}
\noindent
To evaluate the performance of the analytical neighborhood enrichment test, we compared it to the original Monte Carlo-based approach implemented in Squidpy~\cite{squidpy}. Our objectives were twofold: (i) to assess whether the two methods yield similar results by comparing their z-scores, and (ii) to quantify the computational speed-up achieved by the analytical method.

We analyzed eight spatial omics datasets (Table~\ref{table:datasets}) using $k$-nearest neighbor graphs with $k = 6$, 12, and $18$ to define the neighborhoods. For each dataset, we ran the Monte Carlo-based method with $N_{\mathrm{MC}} = 2^i$ label permutations, where $i = 3, \ldots, 10$ (i.e., $N_{\mathrm{MC}}$ ranging from 8 to 1024). Using both methods, we computed z-scores for all label pairs and assessed their agreement via Pearson correlation. To account for the stochastic nature of the Monte Carlo approach, we averaged results over five independent runs. Figure~\figref{fig:results}{A} shows the mean correlation, with error bars indicating the range (max–min) across repetitions. In all datasets, the correlation seemed to plateau after $N_{\mathrm{MC}} = 128$, and exceeded 0.95 in all cases, indicating a strong agreement between the methods.

Next, to assess computational efficiency, we measured the speed-up factor:
\begin{equation*}
\text{Speed-up} = \frac{T_\text{Monte Carlo}}{T_\text{Analytical}},
\end{equation*}
where $T_\text{Monte Carlo}$ and $T_\text{Analytical}$ denote the runtimes of the respective methods. As before, speed-ups were averaged over five repetitions, with ranges shown as error bars. Figure~\figref{fig:results}{B} illustrates how the speed-up increases with $N_{\mathrm{MC}}$, as expected, and tends to be higher for larger datasets (e.g., MERFISH II) and smaller ones, where overhead dominates.

To further explore the relationship between speed-up and dataset size, we fixed $N_{\mathrm{MC}} = 128$, after which the Monte Carlo results had stabilized (as noted in Figure~\figref{fig:results}{A}), and computed the speed-up for $k = 6$. Figure~\ref{fig:results}C shows the resulting speed-up as a function of dataset size, confirming that the speed-up is greater for large datasets.

Lastly, we noted that the z-scores given by the analytical method were slightly lower than those produced by the Monte Carlo-based method, see Figure~\figref{fig:results}{D}.

\subsection*{Analytical neighborhood enrichment score on Xenium data}
\noindent
Finally, we tested the scalability of the analytical neighborhood enrichment score on a large-scale Xenium dataset. We used the {\tt Fresh Frozen Mouse Brain for Xenium Explorer Demo}, publicly available at~\url{https://www.10xgenomics.com/datasets}. This dataset contains 248 mRNA labels and over 40 million spatial points. Using a $k = 6$ nearest neighbor graph and running the computation on a machine with an AMD Ryzen 7 7840HS processor with Radeon 780M Graphics (3.80 GHz) and 16~GB of RAM, the analytical enrichment computation took approximately 90 seconds (excluding graph construction time).

\section{Discussion}
\noindent In this short work, we proposed an analytical version of the neighborhood enrichment score, providing a fast alternative to the Monte Carlo-based method. Our approach yields enrichment scores that are highly correlated with those from the original Monte Carlo sampling method while offering substantial computational speed-up.

Both the Monte Carlo and analytical methods have time complexity proportional to the number of edges in the neighborhood graph. The Monte Carlo method traverses all $E$ edges and checks the labels of connected points. This process is repeated for each of the $N_{\text{MC}}$ label permutations, giving a time complexity of $\mathcal{O}(EN_{\text{MC}})$. The analytical method is dominated by the sparse matrix product $\mathbf{W}\mathbf{L}$, which also has complexity $\mathcal{O}(E)$ since each non-zero in $\mathbf{W}$ touches a single non-zero in $\mathbf{L}$ exactly once.

If $N_{\text{MC}}$ is fixed, this suggests a constant speed-up of the analytical method. However, as shown in Figure~\figref{fig:results}{C}, the observed speed-up increases with dataset size. This is likely due to implementation differences: the Monte Carlo method performs sequential edge traversals in Python (optimized with Numba's just-in-time compiler for performance), while the analytical method uses highly optimized sparse matrix operations (SciPy). For small datasets, the overhead of constructing sparse matrices can offset these gains.

A 70× speed-up was observed for one of the larger datasets (MERFISH II). Note that we set $N_{\text{MC}} = 128$ permutations for the Monte Carlo-based method, while the default in~\cite{squidpy} is $N_{\text{MC}} = 1000$. The Monte Carlo-based method is relatively efficient because the same set of permutations can be used to estimate the null distribution for all label pairs. By contrast, methods like~Ref. \cite{behanova2023visualization} uses a different shuffling strategy: the reference label is fixed, and only the neighbor label is permuted. This requires $N_{\text{MC}}$ permutations for each label pair, increasing computation time. Although not explored here, such alternative null models could be incorporated into the analytical method by modifying the expectation $\mathbb{E}\big[\mathbf{y}\big]$ to a weighted form.

While the analytical enrichment scores are highly correlated with those from the Monte Carlo method in~\cite{squidpy}, they tend to be slightly lower (Figure~\figref{fig:results}{D}). This difference stems from how label randomization is handled: the Monte Carlo method permutes labels without replacement, preserving label counts, whereas the analytical method samples labels independently based on their observed frequencies (effectively sampling with replacement). As a result, the analytical approach does not constrain label counts, likely leading to higher variance in the null distribution (i.e., larger $\sigma_{\text{{\tt AB}}}$), which in turn lowers the resulting z-scores.

The method depends on a single parameter that defines the neighborhood scale, such as $k$ in a $k$-nearest neighbor graph or a radius in a distance-based graph. This, in combination with the computational efficiency, makes the method easy to use for exploring and quantifying spatial colocalization in the data. However, the spatial distributions of the labels in the omics data are rarely completely random. Different biological structures lead to spatial autocorrelation. Thus, comparing the observed counts of neighboring labels against randomly distributed labels tends to result in significant z-scores. As such, scores are best interpreted as relative indicators of spatial association rather than exact measures of statistical significance. 

\bibliographystyle{unsrt}
\bibliography{references}

\end{document}